\documentstyle[preprint,aps,epsf]{revtex}

\newcommand{\la}{\langle}
\newcommand{\ra}{\rangle}
\newcommand{\beq}{\begin{eqnarray}}
\newcommand{\eeq}{\end{eqnarray}}

\newcommand{\btem}{\bibitem}

\begin{document}

\preprint{UTHEP-339, August 1996}

\draft

\title{Tensor Charge of the Nucleon in Lattice QCD}

\author{S. \ Aoki,$^1$
 M.\ Doui$^{1,}$\cite{newaddress},
 T.\ Hatsuda$^1$ and  Y. Kuramashi$^2$}

\address{$^1$Institute of Physics, University of Tsukuba,
 Tsukuba, Ibaraki 305, Japan}

\address{$^2$National Laboratory for High Energy Physics (KEK),
 Tsukuba, Ibaraki 305, Japan}

 
\maketitle

\begin{abstract}

 First results of lattice QCD simulation on the nucleon tensor-charge
  $\delta q$ are presented.
 From the quenched QCD simulations with the Wilson quark action
 at $\beta = 5.7$ on a 16$^3\times$ 20 lattice and
  on a  $12^3\times 20$ lattice,
 we find 
 (i) the connected part $\delta q_{con.}$ 
 is determined with small statistical error, 
  (ii)  the 
  disconnected part $\delta q_{dis.}$, which violates the OZI rule, is
  consistent with zero,
 and (iii) the flavor-singlet tensor
  charge $ \delta \Sigma (\mu^2 = 2 {\rm GeV}^2)
 = \delta u + \delta d + \delta s$
  takes  0.562(88)
  at $\beta=5.7$, which is 
 in contrast with the flavor-singlet axial charge $\Delta \Sigma =
0.1-0.3$.

\end{abstract}

\narrowtext

  The parton structure of the nucleon at the twist two level is known to be
 characterized by three structure functions 
 $f_1(x,\mu)$, $g_1(x,\mu)$ and $h_1(x,\mu)$
 with $x$ being the Bjorken variable and $\mu$ being the renormalization
 scale (see, \cite{JJ}). 
 $f_1$ and $g_1$, which represent the quark-momentum distribution 
 and the quark-spin distribution in the nucleon  respectively, 
 can be measured by the deep inelastic lepton-hadron
 scattering.  On the other hand, $h_1(x)$, which represents  
 the quark-transversity distribution, could be measured in the
 polarized Drell-Yan processes, since it is related to the 
 matrix element of the chiral-odd quark operator.
 Such experiment  is planned using the
  relativistic heavy ion collider (RHIC)
  at BNL. Therefore, theoretical prediction
 of $h_1$ is of great importance. Also, whether there is a 
  large OZI violation in the 
   first moment of $h_1(x)$ has particular interest since 
  the first moment of  $g_1(x)$ and that of the twist three structure
 function $e(x)$ have  been argued to have large OZI violation from
 experimental and/or theoretical point of view \cite{review1,review2,review3}.
   In this paper, we will report first
 results on 
 the  first moment of $h_1(x)$   in  lattice QCD simulations \cite{ADH}.

The tensor charge
 of the nucleon is defined as
$\delta q (\mu) = \int_0^1 [h_1(x,\mu) - \bar{h}_1(x,\mu)]dx$,
 where $h_1(x)$ and $\bar{h}_1(x)$ are the quark and anti-quark
 distributions respectively. 
  One can also write  $\delta q$  as a matrix element of the
 tensor operator
\beq
\label{def2}
\la ps \mid \bar{q} i\sigma_{\mu \nu} \gamma_5 q \mid ps \ra  =  
 2 (s_{\mu} p_{\nu} - s_{\nu} p_{\mu})\  \delta q , 
\eeq
where $p_{\mu}$ is the nucleon's four momentum,
 and $s_{\mu}$ is the nucleon's covariant spin-vector normalized
 as $s_{\mu}^2 = -1$ \cite{notation}.
  In the nucleon light-cone frame, $\delta q$ is interpreted as the
 ``transversity'' of quarks in a transversely polarized nucleon \cite{JJ}.
 On the other hand,
  in the nucleon  rest frame with $\mu =0$ and  $\nu =3$,
 $\delta q \sim \la ps \mid \bar{q} 
  \left( \begin{array}{cc}  \sigma_3 & 0 \\ 0 & \sigma_3 \end{array}
    \right) q  \mid ps \ra $. 
 This implies that $\delta q$ gives an independent information on the 
 spin structure of the nucleon from
 the axial charge 
 $\Delta q \sim
 \la ps \mid \bar{q} \gamma_3 \gamma_5  q \mid ps \ra
 \sim \la ps \mid \bar{q}
  \left( \begin{array}{cc}  \sigma_3 & 0 \\ 0 & - \sigma_3 \end{array}
    \right) q \mid ps \ra $.

 As for $\delta q$,  approximate but model-independent inequalities such as 
 $\mid \delta u \mid < 3/2$ and $\mid \delta d \mid < 1/3$ 
 are known \cite{soff}.
 The non-relativistic quark model predicts
 $ \delta u  =  \Delta u = 4/3$ and 
 $ \delta d  =  \Delta d = -1/3$, while the relativistic quark wave functions
  with non-vanishing lower components
 lead to $\delta u = 1.17$ and $\delta d = -0.29$ together with the
 inequality
 $\mid \delta q \mid > \mid \Delta q \mid $ \cite{HJ}.
  There also exist estimates of $\delta q$ using  QCD sum rules
 \cite{HJ,ioffe}
 and a chiral quark model \cite{goeke}, but the uncertainties
of the results are either large or uncontrollable.

 In lattice QCD simulations,  
  one can treat both
 the connected (OZI preserviong) and disconnected (OZI violating) 
 contributions  even in the quenched approximation
 as has been demonstrated in \cite{kek,liu}.
 Also, one can estimate systematic errors due to  the
 lattice approximation by combining simulations with different
 lattice spacing and/or lattice size.
 
 On the lattice, the matrix element of the tensor operator 
 $O(t,{\bf x}) 
    = \bar{q}  \left( \begin{array}{cc} 
 \sigma_3 & 0 \\ 0 & \sigma_3 \end{array}
    \right)  q $ 
 is extracted as
\beq\
\label{ratio}
 R(t) & \equiv& {\sum_{s = \pm} s
 \la N_s(t) \sum_{t'\neq 0,x} O(t',{\bf x}) \bar{N}_s(0) \ra \over 
\sum_{s = \pm} \la N_s(t) \bar{N}_s(0) \ra } \nonumber \\
& \rightarrow & {\rm const.}
+ Z_T^{-1} \delta q \  t \  \ ({\rm large \ t}),
\eeq
where $Z_T$ is the lattice renormalization factor
 for the tensor operator and  $N(t)$ is 
 the nucleon operator projected to zero momentum.

  Following the works in ref.\cite{kek} where similar simulations have
 been done, we have taken into account the points (i)$\sim$(iv) below.

\vspace{0.2cm}
\noindent
(i)
 We adopt the Wilson quark action and take $ N =
 (q^t C^{-1} \gamma_5 q)q$ as a nucleon interpolating field. 
 To enhance the nucleon signal,
 we use a wall source at initial time slice $t=0$ with the 
   Coulomb gauge fixing at $t=0$. Other time slices are not
 gauge-fixed.
  
\vspace{0.2cm}
\noindent
(ii)
 To avoid the contribution of the negative-parity
 nucleon propagating backward in time, the Dirichlet boundary condition
 for the quark propagator in the temporal direction is adopted.

\vspace{0.2cm}
\noindent
(iii)
 The connected amplitude is calculated by the conventional source
 method \cite{connected}.
  To obtain the disconnected amplitude, the quark propagator
 with space-time unit source without gauge fixing is adopted.
 By this procedure,
 only the gauge invariant closed-loop is automatically selected
  after the average over the gauge configurations \cite{kek}.
 Note that $t'=0$ must be excluded in the sum in eq.(\ref{ratio})
 to make the procedure consistent.

\vspace{0.2cm}
\noindent
(iv)
 To compare the  matrix element on the lattice with that
 in the $\overline{MS}$ scheme, we employ the tadpole-improved
 renormalization factor $Z_T(\mu a)$.
 At  $\mu =1/a$ with $a$ being the lattice spacing,
  $Z_T$ in the Wilson quark action  reads \cite{LM}
\beq
\label{Zfac}
 Z_T = \left[  1 - ({2 \over 3\pi} \ln (\mu a) + 0.44)\  
  \alpha_{\overline{MS}}(1/a) \right]  
 \left( 1- {3 K \over 4 K_c} \right) ,
\eeq
where $K $ is the hopping parameter. The strong coupling constant
$\alpha_{\overline{MS}}(1/a)$ is evolved down by the 2-loop renormalization
 group   from $\alpha_{\overline{MS}}(\pi/a)$ given by
 $\alpha_{\overline{MS}}(\pi/a)^{-1}= P\alpha_0^{-1}+0.30928 $. 
 Here $P$ is the plaquette expectation value
and $\alpha_0$ is the bare lattice coupling constant \cite{alpha}.

\vspace{0.2cm}
We carry out the quenched QCD simulation with the Wilson quark  action
at $\beta =5.7$ on a $16^3 \times 20$ lattice.
In order to estimate the finite size effect,
  simulations 
 at $\beta=5.7$ on a 12$^3 \times 20$ lattice are also performed.
Simulation parameters and some of the 
 measured quantities are summarized in 
Table \ref{tab:tab1}.
Gauge configurations, which are generated with a single plaquette action
 separated by 1000 pseudo heat bath sweeps, are analyzed for
 three hopping parameters. The $u$ and $d$ quarks are assumed to be degenerate
with $K_u=K_d=K_q$, while the strange quark is assigned a different
hopping parameter $K_s$.
  The statistical errors of the data points 
 are estimated by the single elimination 
 jackknife procedure, and final fit of the hadron masses and 
   $R(t)$ are done by the $\chi^2$-fitting.
The lattice spacing $a$ is extracted from $m_\rho a$ in the chiral
limit.

The connected and disconnected parts of $R(t)$ at $\beta=5.7$ on
a $16^3\times 20$ lattice are shown in Fig.~\ref{fig:fig1} at $K=0.164$.
  Clear linear behaviors are seen
 for the u and d connected contributions $\delta q_{con.}$ 
 in the interval up to $ t \sim 11$  even for the smallest
 quark mass $K=0.1665$. For $t > 11$,
  errors grow rapidly.
  The disconnected contribution  $ \delta q_{dis.}$  
   are concentrating around zero with
  errors comparable to the signal below $t=11$.
 Since the disconnected part of $R(t)$ does not show  clear
 signal of a linear slope 
 and we find no other sensible criterion,
 we take the same interval to extract $\delta q_{dis.}$ in our  
 linear fit.  
 We adopt the same interval also for the fit of hadron masses.
 The linear fit of the connected part, the  disconnected part
 and their sum are done separately. 
Our data on $\delta q$ is given in Table \ref{tab:tab2}. 

For u and d quarks the fitted values are linearly extrapolated to the
chiral limit $m_q a =(1/K_q-1/K_c)/2 =0$.
For the strange disconnected contribution, we first make linear interpolation 
to the physical strange quark mass $m_s a=(1/K_s-1/K_c)/2$ at each 
fixed value of $K_q$ for nucleon, 
and then the results are again linearly extrapolated to 
the chiral limit $m_q a =(1/K_q-1/K_c)/2 =0$.
The physical strange quark mass $m_s a$ is estimated
by generalizing the relation $m_{\pi}^2 a^2 = A m_q a$,
obtained from the hadron mass results, to
$m_K^2 a^2 = A (m_q a + m_s a)/2$ and using the experimental ratio
$m_K/m_\rho$=0.64.
The errors of the sum of $u, d$ and $ s$ contributions are estimated by
quadrature.

In Fig.~\ref{fig:fig2}, 
$\delta u_{con.}$, $\delta d_{con.}$ and $\delta u_{dis.}
 = \delta d_{dis.}$ at $\beta=5.7$ on a $16^3\times 20$ lattice are
 shown as a function of $m_qa=(1/K_q -1/K_c)/2$.
 The values linearly extrapolated down to the chiral limit are also
 shown with error bars.

Our final results extrapolated to the chiral limit at
 $\beta=5.7$ on a $16^3\times 20$ lattice  
 are summarized in Table \ref{tab:tab3}.
 The results for $L=12\  (\beta=5.7$), 
 which are done to estimate the finite size effect,
 are also shown in the table.
 To compare our data on $\delta q$ 
 with that of $\Delta q$,
  results of a  previous simulation for $\Delta q$ in ref.\cite{kek} 
 at $\beta =5.7$ on a $16^3 \times 20$ lattice are presented. 
 
 From Table \ref{tab:tab2} and \ref{tab:tab3},
 one finds the following features.


\vspace{0.2cm}
\noindent
 1. Both the connected and total tensor charges 
 at $\beta=5.7$ on a $16^3\times 20$
 lattice satisfy the inequalities 
 $\mid \delta u \mid > \mid \Delta u \mid $ and
 $\mid \delta d \mid < \mid \Delta d \mid $.
 This is different from the prediction of naive quark models
  $ \mid \delta q \mid >
 \mid \Delta q \mid$ mentioned before.

\vspace{0.2cm}
\noindent
2.  We did not see clear linear slope for the disconnected parts
 and we could not make 
  definite conclusion of its precise value from our simulation.
 Nevertheless, rather small error bars ($\sim 0.05$) for $\delta q_{dis.}$
 in our main simulation ($L=16$ at $\beta =5.7$)
 indicate that the OZI violation in $\delta q$ is small with a
  conservative upper bound $\mid \delta q_{dis.} \mid < 0.1$.
 Also, one notices that the disconnected part suffers a large 
 finite volume effect. This is  seen by comparing the data with 
 smaller lattice sizes ($L$=12 at $\beta=5.7$).
 The smallness of $\delta q_{dis.}$ 
  could be related to the C (charge conjugation)-odd and
 chiral-odd  nature
 of the tensor operator $\bar{q} \sigma_{\mu \nu} \gamma_5 q$ \cite{Ji}.

\vspace{0.2cm}
\noindent
3. Due to the small disconnected contributions,
  the flavor singlet tensor charge $\delta \Sigma = \delta u
 + \delta d + \delta s$ is not much suppressed from its
 quark model value. This is in contrast to the
 flavor singlet axial charge $\Delta \Sigma$ which
 is known to be largely suppressed \cite{review1,review3}.

 We have also done the simulation at $\beta=6.0$ on a $16^3 \times 20$ 
 lattice ($a^{-1}=2.271(9) {\rm GeV}$, $La=1.39\  {\rm fm}$)
 and obtained qualitatively the same
  results as above.    
 In the future,  simulations on a larger volume at $\beta=6.0$
 should be done to extract a  definite value of the connected and 
 disconnected tensor charges in the continuum limit as well as to  
 reduce the finite size errors.

\section*{Acknowledgements}

 The authors thank M. Okawa and A. Ukawa
 for helpful discussions and encouragements.
Numerical calculation of this work was done Fujitsu VPP500
 at Univ. of Tsukuba.
 This work was supported in part by 
  Univ. of Tsukuba Research Projects and in part by 
  the Grants-in-Aid of the Japanese Ministry of 
Education, Science and Culture (No. 06102004).


\begin{figure}
\caption{
 $R(t)$ as a function of $t$ at $K=0.164$ and $\beta=5.7$ on a $16^3\times 
 20$ lattice.}
\label{fig:fig1} 
\end{figure}

\begin{figure}
\caption{Quark mass dependence of the tensor charges at
 $\beta=5.7$ on a $16^3\times 
 20$ lattice. Crosses with error bars denote the values extrapolated to
 the chiral limit.}
\label{fig:fig2}
\end{figure}

\newpage

\begin{table}
\begin{center}
\begin{tabular}{|c|ll|}   
$\beta$        &   5.7 &  \\
$L^3\times T$  & $16^3\times 20 $& $12^3 \times 20$ \\ 
 \hline
\# of conf.    & 1053 & 306  \\
$K$            & 0.16, \ 0.164, \ 0.1665 
               & 0.16, \ 0.164, \ 0.1665 \\
fit range      & $6 \leq t \leq 11$ 
               & $5 \leq t \leq 10$  \\
 \hline
$K_c$          & 0.16925(3) & 0.16928(7)  \\
$a^{-1}$ (GeV) & 1.418(9) & 1.452(25)  \\
$La$(fm)       & 2.23 &   1.63  \\
$\alpha_{\overline{MS}}(1/a)$ 
               & 0.2158 & 0.2158  \\ 
 \hline
 $m_\rho a$  & 0.5429(36) & 0.5303(92)   \\
 $m_{_N} a$  & 0.8024(65) & 0.7985(144)  \\
 $m_s a$     & 0.0896(12) & 0.0857(30)   \\
 $ A   $     & 2.752(13)  & 2.746(33)    \\
\end{tabular}
\end{center}
\caption{
Summary of simulation parameters.}
\label{tab:tab1}
\end{table}

\mediumtext 
\begin{table}  
\begin{center}
\begin{tabular}{|ccccccc|}   
$K_q$ & $\delta u_{con.}$ & $\delta d_{con.}$ &
$\delta u_{dis.}=\delta d_{dis.}$ & \multicolumn{3}{c|}{$\delta s_{dis.}$\ } \\
\hline
\multicolumn{4}{|l}{$\beta=5.7$, $16^3\times 20$} & $K_s$=0.1665 & 0.164 
& 0.160 \\
0.1665& 0.948(31) & -0.196(15) & -0.044(69) & -0.044(69) &-0.035(58)
& -0.009(51) \\
0.164 & 0.994(11) & -0.220(52) & -0.027(28) & -0.041(33) &-0.027(28)
& -0.013(25) \\
0.160 & 1.072(50) & -0.251(21) & -0.008(12) & -0.022(16) &-0.013(14)
&-0.008(12) \\
\hline
 \multicolumn{4}{|l}{$\beta=5.7$, $12^3\times 20$} & $K_s$=0.1665 & 0.164
& 0.160 \\
0.1665 & 0.825(49) & -0.227(23)& 0.081(84)& 0.081(85)&0.067(74)&0.053(66)\\
0.164  & 0.919(20) & -0.242(9) & 0.034(38)& 0.035(44)&0.034(38)&0.026(34)\\
0.160  & 1.034(10) & -0.257(4) & 0.014(17)& 0.016(22)&0.018(19)&0.014(17)\\
\end{tabular}
\end{center}
\caption{Measured tensor charges at $\mu = 1/a$.}
\label{tab:tab2}
\end{table}

\begin{table}  
\begin{center}
\begin{tabular}{|lll|ll|}
\multicolumn{3}{|l|}{tensor charge} &
\multicolumn{2}{l|}{axial charge (ref.\cite{kek})} \\
\hline
$\beta$ &\multicolumn{2}{l|}{5.7}  & $\beta$ & 5.7 \\
size & $16^3\times 20$&$12^3\times 20$ & 
size & $16^3\times 20$ \\
$\delta u$        
   &   0.839(60)\  &  0.822(83)\    &$\Delta u$& 0.638(54) \\
 $\delta u_{con.}$ 
   &   0.893(22)\  &  0.760(39)\    &$\Delta u_{con.}$&0.763(35)\\
 $\delta d$   
   & -0.231(55)\   &  -0.159(75)\   &$\Delta d$&-0.347(46) \\
 $\delta d_{con.}$ 
   & -0.180(10)\   &  -0.220(17)\   &$\Delta d_{con.}$&-0.226(17) \\
 $\delta (u,d)_{dis.}$ 
 & -0.054(54)    & 0.076(71)\       &$\Delta (u,d)_{dis.}$&-0.119(44)
\\
 $\delta s_{dis.}$
   & -0.046(34)    & 0.071(46)\     &$\Delta s_{dis.}$&-0.109(30)
 \\ 
\hline
$\delta \Sigma $ 
   & 0.562(88)\    &  0.733(121)      
&$\Delta \Sigma $& 0.18(10)\\
 \end{tabular}
\end{center}
\caption{Tensor charges extrapolated to the chiral limit
together with the axial charge obtained in ref.\protect{\cite{kek}}.
 The flavor singlet tensor charge is defined as
 $\delta \Sigma = \delta u + \delta d + \delta s $.
 The matrix elements are evaluated at $\mu = 1/a$.}
\label{tab:tab3}
\end{table}

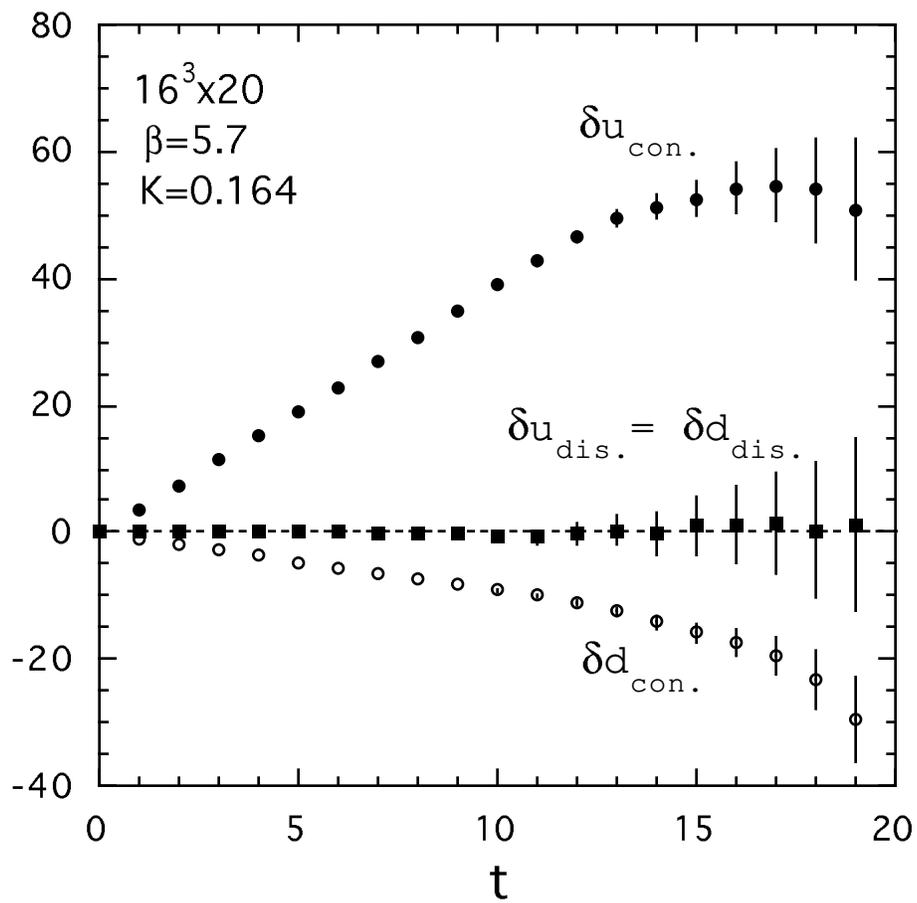
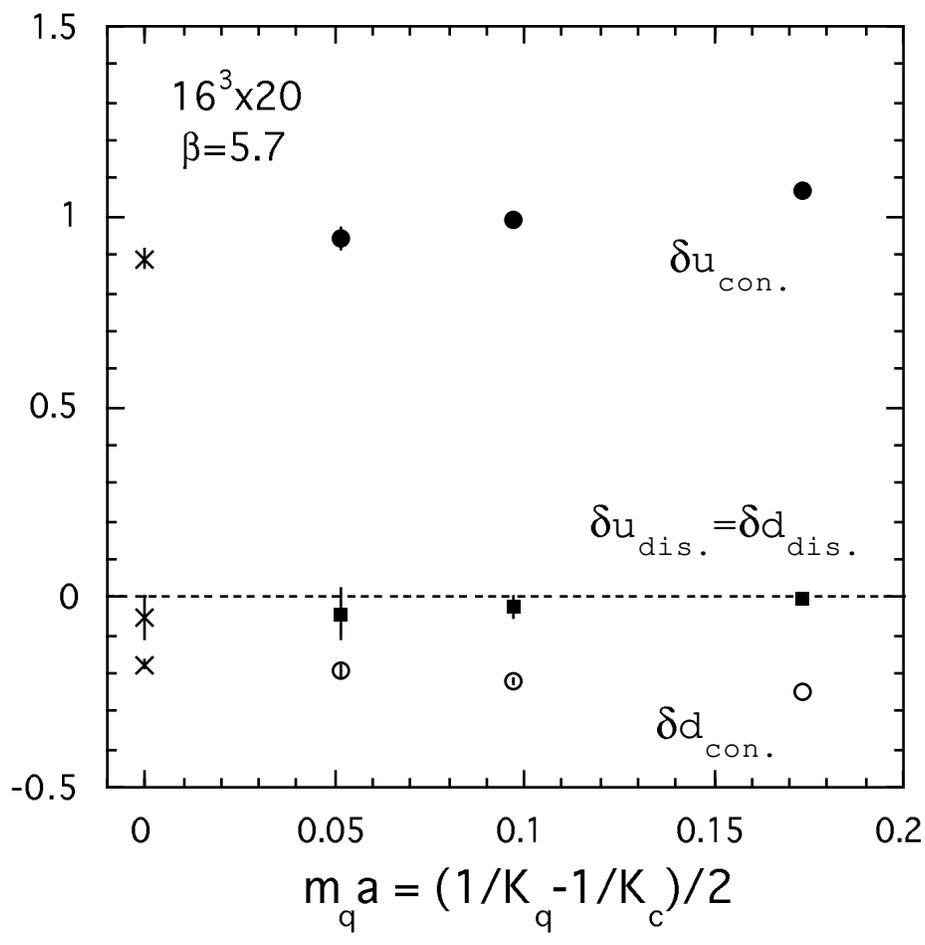
\end{document}